\documentclass[twocolumn, pra]{revtex4}
\usepackage{graphicx}
\usepackage{dcolumn}
\usepackage{bm}
\usepackage{textcomp}
\usepackage{amssymb}
\usepackage{mathrsfs}
\usepackage{amsmath}
\allowdisplaybreaks[4]
\usepackage[colorlinks, linkcolor=blue, anchorcolor=blue,citecolor=blue]{hyperref}

\begin{document}
\title{Ultrastrong-coupling quantum-phase-transition phenomena in a few-qubit circuit QED system}

\author{Wen-Juan Yang$ ^{1, 2}$, Xiang-Bin Wang$ ^{1, 2, 3\footnote{Email
address: xbwang@mail.tsinghua.edu.cn}}$}

\affiliation{ \centerline{$^{1}$State Key Laboratory of Low
Dimensional Quantum Physics, Tsinghua University, Beijing 100084,
People's Republic of China}\centerline{$^{2}$ CAS Center for Excellence and Synergetic Innovation Center in Quantum Information and Quantum Physics,}
 \centerline{University of Science and Technology of China, Hefei, Anhui 230026, China}\centerline{$^{3}$ Jinan Institute of Quantum Technology, Jinan, Shandong 250101, China}}

\begin{abstract}
 We study ultrastrong-coupling quantum-phase-transition phenomena in a few-qubit system. In the one-qubit case, three second-order transitions occur and the Goldstone mode emerges under the condition of ultrastrong-coupling strength. Moreover, a first-order phase transition occurs between two different superradiant phases. In the two-qubit case, a two-qubit Hamiltonian with qubit-qubit interactions is analyzed fully quantum mechanically. We show that the quantum phase transition is inhibited even in the ultrastrong-coupling regime in this model. In addition, in the three-qubit model, the superradiant quantum phase transition is retrieved in the ultrastrong-coupling regime. Furthermore, the $N$-qubit model with $U(1)$ symmetry is studied and we find that the superradiant phase transition is inhibited or restored with the qubit-number parity.
\end{abstract}

\maketitle
\parskip=0 pt

\section{Introduction}
The Dicke model, where a large number of atoms interact with a single radiation mode, has attracted much attention in the studies of quantum electrodynamics (QED) in the cavity and superconductor circuit systems since it was first proposed by Dicke\cite{dicke1, dicke2, cavity1, cavity2, cavity3, cavity4, cavity5, cavity6, circuit1}. It was later found that in the thermodynamical limit, i.e., the atom number $N\rightarrow\infty$, and in the strong-coupling regime, the model exhibits a superradiant quantum phase transition (QPT)\cite{pt1, pt2, pt3, pt4, pt5, pt6, qubit}. In a large $N$ limit and a single-qubit coupling strength beyond the critical point, the effective coupling between atoms and the radiation mode becomes comparable to the bare frequencies of the  atom and radiation mode, which leads to the occurrence of a superradiant phase transition.

Recently, a similar QPT behavior has been demonstrated in the limit where the ratio ($\omega_q/\omega_r$) of the qubit frequency to the single-mode resonator frequency diverges\cite{flimit1, flimit2, flimit3, flimit4, flimit5, flimit6, flimit7, flimit8}. Meanwhile, the experimental realization of an ultrastrong-coupling regime for single-qubit and single-mode resonators in superconductor circuit systems has been realized\cite{exp1, exp2, exp3, exp4}. This makes it possible to investigate the QPT phenomena in a single-qubit or few-qubit level in a controllable circuit system.

In this paper, first, we consider a superconductor configuration where a single qubit is coupled both inductively and capacitively to a resonator which induces two different kinds of atom-resonator coupling terms $g_x$ and $g_y$, respectively. As the Dicke Hamiltonian with only one kind of coupling term has a discrete $Z_2$ symmetry, the particle non-conserving terms cannot be neglected in the ultrastrong-coupling regime and the Hamiltonian does not have $U(1)$ symmetry. However, in our work, for the Hamiltonian with two different kinds of coupling terms $g_x$ and $g_y$, the continuous $U(1)$ symmetry is preserved for $g_x=g_y$. For such a system, we find that three different superradiant phases can occur in the situation where $\omega_q/\omega_r\rightarrow\infty$ and $g_x$ or $g_y$ beyond the critical points. In particular, in the superradiant phase where $g_x=g_y$, the Goldstone mode emerges. Moreover, a first-order phase transition occurs between two different superradiant phases when two different discrete symmetries are broken. Our analysis is in agreement with the study proposed in Ref. \cite{pt6} in the thermodynamical limit where $N\rightarrow\infty$. Second, we consider a circuit configuration where two atoms are coupled to the same cell resonator in a transmission line, both inductively and capacitively, and qubit-qubit interactions are involved. We find that in this model, the quantum phase transition is inhibited even in the ultrastrong-coupling regime. Third, we consider a model with three qubits. We find that even in the presence of qubit-qubit coupling, a superradiant phase transition is retrieved. Fourth, an effective Hamiltonian for $N$ qubits is given. For such a system, we find that the presence of a superradiant phase transition depends on the qubit-number parity.

The paper is organized as follows. In Sec. II we propose a one-qubit circuit QED system. We derive the effective low-energy Hamiltonian and discuss the phases of the one-qubit model. In Sec. III we consider a two-qubit circuit QED system with qubit-qubit coupling. We derive the effective low-energy Hamiltonian and discuss the phases of the two-qubit model. In Secs. IV and V, we extend the situation to three-qubit and $N$-qubit situations. Finally, Sec. VI summarizes the
main conclusions of this work.
\section{One-Qubit Model}
\begin{figure}
  \centering
  \includegraphics[width=0.45\textwidth]{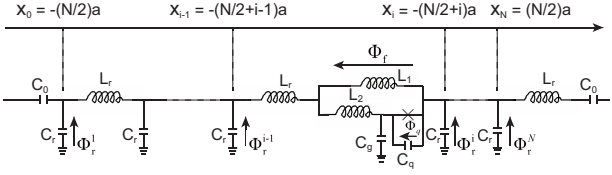}\\
  \caption{Superconducting circuit model with one atom coupled both inductively and capacitively to a transmission line resonator. }\label{fig1}
\end{figure}

Figure \ref{fig1} shows a superconducting circuit with one artificial atom.
 The artificial atom is coupled both inductively and capacitively (with capacitance $C_g$) to a transmission line resonator with inductance ${L}_r$ and capacitance ${C}_r$. The atom (fluxonium) consists of a Josephson junction with capacitance ${C}_q$ and Josephson energy ${E}_J$ coupled to inductances $L_1$ and $L_2$\cite{pt6,patent}. The Lagrangian of the circuit reads
\begin{eqnarray}
\mathcal{L}&=&C_r\frac{(\dot{\Phi}_r^i)^2}{2}+C_q\frac{(\dot{\Phi}_q)^2}{2}+E_J\cos{(\frac{\Phi_q+\Phi_{ext}}{\Phi_0})} \nonumber\\
&&+C_g\frac{(\dot{\Phi}_r^i+
\dot{\Phi}_q)^2}{2}
-\frac{\Phi_f^2}{2L_1}-\frac{(\Phi_f-\Phi_q)^2}{2L_2}\nonumber\\
&&-\frac{(\Phi_r^{i-1}-
\Phi_r^i-\Phi_f)^2}{2L_r}, \label{L1}
\end{eqnarray}
where $\Phi_0=\hbar/(2e)$ is the flux quantum and $\Phi_{\mathrm{ext}}$ is the external flux.
By definition, the charge $\mathrm{Q}_i$ is conjugate to the flux $\Phi_i$ (obeying $[\Phi_i, \mathrm{Q}_j]=i\hbar\delta_{ij}$), and hence $\mathrm{Q}_i=\partial\mathcal{L}/\partial\dot{\Phi}_i$:
\begin{eqnarray}
\left(\begin{array}{*{20}{c}}
\mathrm{Q}_r^i\\
\mathrm{Q}_q
\end{array}\right)&=&\left(\begin{array}{cc}
C_r+C_g&C_g\\
C_g&C_q+C_g\end{array}\right)\left(\begin{array}{*{20}{c}}
\dot{\Phi}_r^i\\
\dot{\Phi}_q\end{array}\right).\label{Q1}
\end{eqnarray}
By applying Kirchoff's law, we can get the relation
\begin{eqnarray}
\frac{\Phi_r^{i-1}-\Phi_r^i-\Phi_f}{L_r}=\frac{\Phi_f}{L_1}+\frac{\Phi_f-\Phi_q}{L_2}.\label{phif}
\end{eqnarray}
By applying Eqs. (\ref{L1})-(\ref{phif}) and the definition $\mathcal{H}=\mathrm{Q}_r\dot{\Phi}_r+\mathrm{Q}_q\dot{\Phi}_q-\mathcal{L}$, we obtain
\begin{eqnarray}
\mathcal{H}&=&\mathcal{H}_r+\mathcal{H}_q+\mathcal{H}_{int},\\
\mathcal{H}_r&=&\frac{(\mathrm{Q}_r^i)^2}{2\bar{C}_r}+\frac{(\Phi_r^{i-1}-\Phi_r^i)^2}{2\bar{L}_r},\\
\mathcal{H}_q&=&\frac{\mathrm{Q}_q^2}{2\bar{C}_q}
+\frac{\Phi_q^2}{2\bar{L}_q}-E_J\cos{(\frac{\Phi_q+\Phi_{ext}}{\Phi_0})},\\
\mathcal{H}_{int}&=&-\frac{\mathrm{Q}_r^i\mathrm{Q}_q}{\bar{C}_g}-\frac{(\Phi_r^{i-1}-\Phi_r^{i})\Phi_q}{\bar{L}_g},
\end{eqnarray}
where $\bar{C}_r=C_{\Sigma}^2/(C_q+C_g)$, $\bar{L}_r=L_{\Sigma}^2/(L_1+L_2)$, $\bar{C}_q=C_{\Sigma}^2/(C_r+C_g)$, $\bar{L}_q=L_{\Sigma}^2/(L_1+L_r)$, $\bar{C}_g=C_{\Sigma}^2/C_g$, $\bar{L}_{g}=L_{\Sigma}^2/L_1$, $C_{\Sigma}^2=C_rC_g+C_rC_q+C_gC_q$, and $L_{\Sigma}^2=L_rL_1+L_rL_2+L_1L_2$.

By quantizing the transmission line resonator mode, we express $\Phi_r^i$ as $\Phi_r(x_i)=\cos(kx_i)\sqrt{\frac{\hbar}{\Omega_rC_r}}(a+a^{\dagger})$ and $\mathrm{Q}_r^i$ as $\mathrm{Q}_r(x_i)=-i\cos(kx_i)\sqrt{\hbar\Omega_rC_r}(a-a^{\dagger})$, where $\Omega_r=\frac{\pi a}{d\sqrt{L_rC_r}}$, $k=\frac{\pi}{d}$, and $d$ is the length of the transmission line resonator. Note that here we only consider the lowest resonant mode of the transmission line resonator. Next, utilizing a two-level system approximation for the qubit $\mathcal{H}_q$ (with $\Phi_{\mathrm{ext}}=\pi\Phi_0$), we can write $H_q=\hbar\omega_q\sigma_z/2$, $\Phi_q=\Phi_0^q\sigma_x$ ($\Phi_0^q=\langle g|\Phi_q|e\rangle$), and $\mathrm{Q}_q=-\omega_q\bar{C}_q\Phi_0^q\sigma_y$, where $|g\rangle$ and $|e\rangle$ are the ground state and first excited state of the qubit Hamiltonian, respectively. Finally, we obtain the following quantized Hamiltonian ($\hbar=1$),
\begin{eqnarray}
\mathcal{H}&=&\omega_rb^\dagger b+\omega_q\frac{\sigma_z}{2}-g_x(b+b^\dagger){\sigma_x\over2}\nonumber\\
&&-ig_y(b-b^\dagger){\sigma_y\over2},\label{Hamiltonian}
\end{eqnarray}
where $g_x=2|\cot(kx_i)|^{{1/2}}\sin(kx_i)\frac{\sqrt{\Omega_rL_r}}{\bar{L}_g}({{C_r\bar{L}_r}\over{\bar{C}_rL_r}})^{1/4}\Phi_0^q$, $g_y=2|\tan(kx_i)|^{{1/2}}\cos(kx_i)\sqrt{\Omega_rC_r}\frac{\bar{C}_q}{\bar{C}_g}({{\bar{C}_rL_r}\over{C_r\bar{L}_r}})^{1/4}\omega_q\Phi_0^q$ and $\omega_r=|\sin(2kx_i)|\Omega_r(\frac{{C}_r{L}_r}{\bar{C}_r\bar{L}_r})^{{1\over 2}}$. The Hamiltonian given by Eq. (\ref{Hamiltonian}) commutes with the parity operator $\Pi=e^{i\pi(b^\dagger b+\sigma_+\sigma_-)}$ and it possesses a discrete $Z_2$ symmetry. Especially, when $g_x=g_y$, the Hamiltonian is reduced to a Jaynes-Cummings (JC) model and it possesses a continuous $U(1)$ symmetry which is invariant under a gauge transformation $e^{i\vartheta(b^\dagger b+\sigma_+\sigma_-)}$.

Here, we consider the parameter regime where $\omega_q\gg\omega_r$. In this limit, we can make a unitary transformation $U=e^S=\exp[
\frac{g_x+g_y}{2\omega_q}(b^\dagger\sigma_{-}-b\sigma_{+})+\frac{g_x-g_y}{2\omega_q}(b\sigma_{-}-b^\dagger\sigma_{+})]$, which is equivalent to the adiabatic elimination. After this transformation, the obtained effective Hamiltonian decouples the low-energy (spin-down) and high-energy (spin-up) spin subspace. Projecting the effective Hamiltonian to the low-energy subspace, we obtain
\begin{eqnarray}
\mathcal{H}_{\mathrm{eff}}&=&P(UHU^\dagger)P=\omega_rb^\dagger b-\omega_r\frac{\lambda_x^2}{4}(b+b^\dagger)^2\nonumber\\
&&+\omega_r\frac{\lambda_y^2}{4}(b-b^\dagger)^2-\frac{\omega_q}{2}, \label{Heff}
\end{eqnarray}
where $\lambda_x=\frac{g_x}{\sqrt{\omega_r\omega_q}}$ and $\lambda_y=\frac{g_y}{\sqrt{\omega_r\omega_q}}$. The detailed derivation is shown in Appendix \ref{appendixa}. Diagonalizing Hamiltonian Eq. (\ref{Heff}), we obtain $\mathcal{H}_{\mathrm{eff}}=\epsilon c^\dagger c-\frac{\omega_q}{2}$ with $\epsilon=\omega_r\eta\sqrt{(1-\lambda_x^2)(1-\lambda_y^2)}$. Here, $\eta=1$ for $\lambda_x^2+\lambda_y^2<2$ and $\eta=-1$ for $\lambda_x^2+\lambda_y^2>2$. Operator $b$ is related to operator $c$ by $c=\hat{S}^\dagger(r)b\hat{S}(r)$, where $\hat{S}(r)=\exp[\frac{1}{2}r(b^2-b^{\dagger2})]$, with $r={1\over 4}\ln\frac{(1-\lambda_y^2)}{(1-\lambda_x^2)}$. It is easy to see that $\epsilon$ is imaginary for $(1-\lambda_x^2)(1-\lambda_y^2)<0$. For $(1-\lambda_x^2)(1-\lambda_y^2)>0$ and $\lambda_x^2+\lambda_y^2>2$, the characteristic energy $\epsilon$ is negative. The above two abnormal phenomena suggest the failure of Eq. (\ref{Heff}) to describe the low-energy property and a higher-
order subspace should be taken into consideration. These phenomena also suggest the occurrence of superradiant phases in the above two cases.

To investigate the low-energy property, we consider the following transformed Hamiltonian,
\begin{eqnarray}
\tilde{\mathcal{H}}&=&\mathcal{D}^\dagger[\alpha]\mathcal{H}\mathcal{D}[\alpha]=\omega_rb^\dagger b+\tilde{\omega}_q\frac{\tau_z}{2}+\omega_r|\alpha|^2\nonumber\\
&&-{g_x\over 2}(b+b^\dagger)(\cos{\phi}\cos{2\theta}\tau_x+\sin{\phi}\tau_y)\nonumber\\
&&-i{{g}_y\over 2}(b-b^\dagger)(-\sin{\phi}\cos{2\theta}\tau_x+\cos{\phi}\tau_y),
\end{eqnarray}
where $\mathcal{D}[\alpha]$ is the displacement operator with $\alpha=\pm{1\over 2}\sqrt{{\omega_q\over\omega_r}(\lambda_x^2-{1\over\lambda_x^2})}$ for $\lambda_x>1$ and $\lambda_x>\lambda_y$, and $\alpha=\pm {i\over 2}\sqrt{{\omega_q\over\omega_r}(\lambda_y^2-{1\over\lambda_y^2})}$ for $\lambda_y>1$ and $\lambda_y>\lambda_x$. Note that the displacement is $\alpha={e^{i\vartheta}\over 2}\sqrt{{\omega_q\over\omega_r}(\lambda_x^2-{1\over\lambda_x^2})}$ $[\vartheta\in[0,2\pi)]$ for $g_x>1$ and $g_x=g_y$ as in Ref. \cite{flimit8}. Here, $\tilde{\omega}_q=\omega_q\lambda_x^2$, $\tau_z=|\tilde{\uparrow}\rangle\langle\tilde{\uparrow}|-|\tilde{\downarrow}\rangle\langle\tilde{\downarrow}|
=\cos{2\theta}\sigma_z+\sin{2\theta}\sigma_x$, $\tan{2\theta}=\mp\frac{2\sqrt{g_x^2|\alpha|}}{\omega_q}$, and $\phi=0$ for $\lambda_x>1$ and $\lambda_x>\lambda_y$, while  $\tilde{\omega}_q=\omega_q\lambda_y^2$, $\tau_z=|\tilde{\uparrow}\rangle\langle\tilde{\uparrow}|-|\tilde{\downarrow}\rangle\langle\tilde{\downarrow}|
=\cos{2\theta}\sigma_z-\sin{2\theta}\sigma_y$, $\tan{2\theta}=\mp\frac{2\sqrt{g_y^2|\alpha|}}{\omega_q}$, and $\phi={\pi\over 2}$ for $\lambda_y>1$ and $\lambda_y>\lambda_x$ . Note that here the qubit frequency and qubit-resonator coupling coefficients are rescaled. Following the same low-energy approximation as that of Eq. (\ref{Hamiltonian}), we get
\begin{eqnarray}
\tilde{\mathcal{H}}_{\mathrm{eff}}&=&\omega_r b^\dagger b-\omega_r\frac{\tilde{\lambda}_x^2}{4}(b+b^\dagger)^2+\omega_r\frac{\tilde{\lambda}_y^2}{4}(b-b^\dagger)^2\nonumber\\
&&-\frac{\tilde{\omega}_q}{2}+\omega_r|\alpha|^2, \label{tildeHeff}
\end{eqnarray}
where $\tilde{\lambda}_x=\frac{1}{\lambda_x^2}$ and $\tilde{\lambda}_y=\frac{\lambda_y}{\lambda_x}$ for $\lambda_x>1$ and $\lambda_x>\lambda_y$, and $\tilde{\lambda}_x=\frac{\lambda_x}{\lambda_y}$ and $\tilde{\lambda}_y=\frac{1}{\lambda_y^2}$ for $\lambda_y>1$ and $\lambda_y>\lambda_x$. By diagonalizing Hamiltonian Eq. (\ref{tildeHeff}), we obtain $\tilde{\mathcal{H}}_{\mathrm{eff}}=\tilde{\epsilon}\tilde{c}^\dagger \tilde{c}-\frac{\tilde{\omega}_q}{2}
+\omega_r|\alpha|^2$, with $\tilde{\epsilon}=\omega_r\sqrt{(1-\tilde{\lambda}_x^2)(1-
\tilde{\lambda}_y^2)}$. Here, $\tilde{c}=\hat{S}^\dagger(\tilde{r})b\hat{S}(\tilde{r})$ with $\hat{S}(\tilde{r})=\exp[\frac{1}{2}\tilde{r}(b^2-b^{\dagger2})]$ and $\tilde{r}={1\over 4}\ln\frac{(1-\tilde{\lambda}_y^2)}{(1-\tilde{\lambda}_x^2)}$. The eigenstates of the system are $|\psi\rangle=D[\alpha]S[\tilde{r}]|m\rangle|\tilde{\downarrow}\rangle$, which implies a nonzero coherence of the resonator field where $\langle a\rangle=\alpha$. Note that in the case $\lambda_x=\lambda_y$ and $\lambda_x>1$, we have $\tilde{\epsilon}=0$. This indicates that the Goldstone mode emerges.

Our analysis shows that the critical point of the superradiant phase transition appears at $\lambda_x=1$ or $\lambda_y=1$. The ground-state resonator-mode number is $n_G=0$ for $\lambda_x<1$ and $\lambda_y<1$, and $n_G=|\alpha|^2$ for $\lambda_x>1$ or $\lambda_y>1$. Therefore, $n_G$ is an order parameter. The ground-state energy is $\epsilon_G=-{\omega_q\over 2}$ for $\lambda_x<1$ and $\lambda_y<1$, $\epsilon_G=-{\omega_q\over 4}(\lambda_x^2+\frac{1}{\lambda_x^2})$ for $\lambda_x>1$ and $\lambda_x>\lambda_y$, and $\epsilon_G=-{\omega_q\over 4}(\lambda_y^2+\frac{1}{\lambda_y^2})$ for $\lambda_y>1$ and $\lambda_y>\lambda_x$. The ground-state energy $\epsilon_G$ is continuous, while $\partial^2\epsilon/\partial^2 g_x$ is discontinuous at $\lambda_x=1$ and $\lambda_x>\lambda_y$, so that the normal-phase to the superradiant-phase transition is of second order. Analogously, $\partial^2\epsilon/\partial^2 g_y$ is discontinuous at $\lambda_y=1$ and $\lambda_y>\lambda_x$, which indicates a second-order phase transition. Moreover, $\partial\epsilon/\partial g_x$ and $\partial\epsilon/\partial g_y$ are discontinuous at $\lambda_x>1$ and $\lambda_x=\lambda_y$, therefore the transition between these two kinds of superradiant phases is of first order. The corresponding phase \begin{figure}
  \centering
  \includegraphics[width=0.4\textwidth]{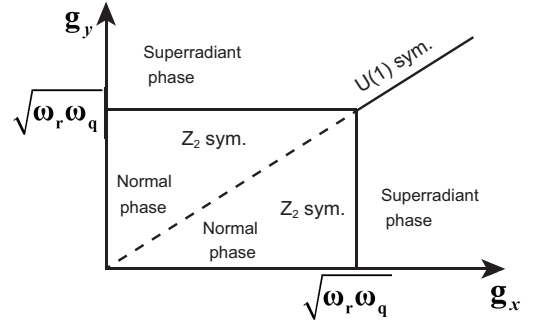}\\
  \caption{Phase diagram of the Hamiltonian Eq. (\ref{Hamiltonian}) in the $(g_x, g_y)$ plane.}\label{fig2}
\end{figure}
diagram is shown in Fig. \ref{fig2}.

\section{Two-Qubit Model}
\begin{figure}
  \centering
  \includegraphics[width=0.5\textwidth]{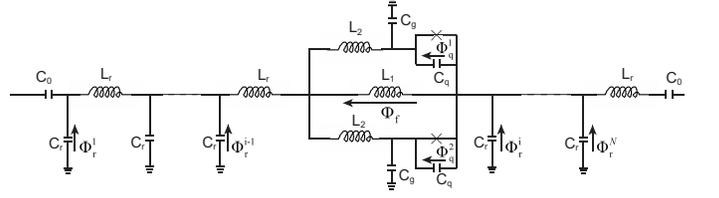}\\
  \caption{Superconducting circuit configuration with two atoms. Two fluxoiums are coupled both inductively and capacitively to the same cell resonator of the transmission line resonator.}\label{fig3}
\end{figure}

The superconducting circuit configuration with two atoms is illustrated in Fig. \ref{fig3}.
Two fluxoiums are coupled both inductively and capacitively to the same cell resonator of the transmission line resonator. The Lagrangian of the circuit reads
\begin{eqnarray}
\mathcal{L}_2&=&C_r\frac{(\dot{\Phi}_r^i)^2}{2}-\frac{\Phi_f^2}{2L_1}-\frac{(\Phi_r^{i-1}-\Phi_r^i-\Phi_f)^2}{2L_r}\nonumber\\
&&+\sum_{j=1}^2 [C_q\frac{(\dot{\Phi}_q^j)^2}{2}+E_J\cos(\frac{\Phi_q^j+\Phi_{ext}^j}{\Phi_0})\nonumber\\
&& +C_g\frac{(\dot{\Phi}_r^i+
\dot{\Phi}_q^j)^2}{2}
-\frac{(\Phi_f-\Phi_q^j)^2}{2L_2}]. \label{L2}
\end{eqnarray}
According to Eq. (\ref{L2}), the corresponding Hamiltonian is
\begin{eqnarray}
\mathcal{H}_2&=&\mathcal{H}_{r}'+\mathcal{H}_q'+\mathcal{H}_{int}',\\
\mathcal{H}_r'&=&\frac{(\mathrm{Q}_r^i)^2}{2\bar{C}_r'}+\frac{(\Phi_r^{i-1}-\Phi_r^i)^2}{2\bar{L}_r'},\\
\mathcal{H}_q'&=&\sum_{j=1}^2[\frac{\mathrm{Q}_q^{j2}}{2\bar{C}_q'}
+\frac{\Phi_q^{j2}}{2\bar{L}_q'}-E_J\cos(\frac{\Phi_q^j+\Phi_{ext}^j}{\Phi_0})],\\
\mathcal{H}_{int}'&=&\sum_{j=1}^2[-\frac{\mathrm{Q}_r^i\mathrm{Q}_q^j}{\bar{C}_g'}-\frac{(\Phi_r^{i-1}-\Phi_r^{i})\Phi_q^j}{\bar{L}_g'}\nonumber\\
&&+\frac{\mathrm{Q}_q^1\mathrm{Q}_q^2}{\bar{C}_{qq}}-\frac{\Phi_q^1\Phi_q^2}{\bar{L}_{qq}}],
\end{eqnarray}
where $\bar{C}_r'=C_{\Sigma}'^2/(C_q+C_g)$, $\bar{L}_r'=L_{\Sigma}'^2/(2L_1+L_2)$, $\bar{C}_q'=C_{\Sigma}'^2/[C_r+C_g(1+\frac{C_q}{C_g+C_q})]$, $\bar{L}_q'=L_{\Sigma}'^2/(L_1+L_r+L_1L_r/L2)$, $\bar{C}_g'=C_{\Sigma}'^2/C_g$, $\bar{L}_{g}'=L_{\Sigma}'^2/L_1$, $\bar{C}_{qq}=C_{\Sigma}'^2/\frac{C_g^2}{C_g+C_q}$, $\bar{L}_{qq}=L_{\Sigma}'^2/\frac{L_1L_r}{L_2}$, $C_{\Sigma}'^{2}=C_rC_g+C_rC_q+2C_gC_q$, and $L_{\Sigma}'^2=2L_rL_1+L_rL_2+L_1L_2$.  Note that qubit-qubit interaction terms are added in the two-qubit Hamiltonian. Following the same quantizing procedure as in the one-qubit case, we obtain the following Hamiltonian $\mathcal{H}_2$,
\begin{eqnarray}
\mathcal{H}_2&=&\omega_r'd^\dagger d+\omega_q'S_z-g_x'(d+d^\dagger)S_x-ig_y'(d-d^\dagger)S_y\nonumber\\
&&-D_xS_x^2+D_yS_y^2, \label{H2}
\end{eqnarray}
where $g_x'=2|\cot(kx_i)|^{{1/2}}\sin(kx_i)\frac{\sqrt{\Omega_rL_r}}{\bar{L}_g'}({{C_r\bar{L}_r'}\over{\bar{C}_r'L_r}})^{1/4}\Phi_0^{q'}$, $g_y'=2|\tan(kx_i)|^{{1/2}}\cos(kx_i)\sqrt{\Omega_rC_r}\frac{\bar{C}_q'}{\bar{C}_g'}({{\bar{C}_r'L_r}\over{C_r\bar{L}_r'}})^{1/4}
\omega_q'\Phi_0^{q'}$, $\omega_r'=|\sin(2kx_i)|\Omega_r(\frac{{C}_r{L}_r}{\bar{C}_r'\bar{L}_r'})^{{1\over 2}}$, and $S_k(k=x,y,z)=\frac{\sigma_1^{k}}{2}+\frac{\sigma_2^{k}}{2}$. The coefficients $D_x$ and $D_y$ are not independent. After calculation, one finds that $D_y$ happens to be $\frac{g_y'^2}{\omega_r'}$ while $D_x=\frac{g_x'^2}{\omega_r'}$ for $L_2=\frac{2L_1L_r}{L_1-L_r}$. The total excitation number $N_2=d^\dagger d+\sigma_1^+\sigma_1^-+\sigma_2^+\sigma_2^-$ is not conserved for $g_x=g_y$. Therefore the continuous $U(1)$ symmetry is not preserved. Since $[e^{i\vartheta(d^\dagger d+\sigma_1^+\sigma_1^-+\sigma_2^+\sigma_2^-)}, S_x^2]=-[e^{i\vartheta(d^\dagger d+\sigma_1^+\sigma_1^-+\sigma_2^+\sigma_2^-)}, S_y^2]$, the $U(1)$ symmetry is restored if the sign of $D_x$ or $D_y$ is reversed. Following the method shown in Refs. \cite{simulation1, simulation2}, the sign of the qubit-qubit coupling coefficients can be reversed by applying a sequence of local qubit rotations. The obtained Hamiltonian is
\begin{eqnarray}
\mathcal{H}_3&=&3\omega_r'd^\dagger d+2\omega_q'S_z-g_x'(d+d^\dagger)S_x\nonumber\\
&&-ig_y'(d-d^\dagger)S_y+D_xS_x^2+D_yS_y^2. \label{H3}
\end{eqnarray}
In order to diagonalize the qubit-qubit coupling terms, spin operators $\sigma_{1,2}^k$ are transformed to fermionic operators via the Jordan-Wigner transformation,
\begin{eqnarray}
f_1&=&\sigma_1^{-},\qquad f_2=-\sigma_1^{z}\sigma_2^{-},\nonumber\\
f_1^\dagger&=&\sigma_1^{+},\qquad f_2^\dagger=-\sigma_1^{z}\sigma_2^{+},
\end{eqnarray}
where $\sigma_i^{+}=\frac{1}{2}(\sigma_i^{x}+i\sigma_i^{y})$, $\sigma_i^{-}=\frac{1}{2}(\sigma_i^{x}-i\sigma_i^{y})$. After a linear transformation of fermionic operators, the Hamiltonian $\mathcal{H}_3^0=2\omega_q'S_z+D_xS_x^2+D_yS_y^2$ can be diagonalized as $\mathcal{H}_3^0=\Lambda_1\eta_1^\dagger\eta_1+\Lambda_2\eta_2^\dagger\eta_2$ with $\Lambda_i(i=1,2)=\frac{1}{2}\omega_q'[\pm(\lambda_x'^2+\lambda_y'^2)+\sqrt{(\lambda_x'^2-\lambda_y'^2)^2+16}]$ for $\lambda_x'^2\lambda_y'^2<4$ and $\Lambda_{1,2}=\frac{1}{2}\omega_q'[(\lambda_x'^2+\lambda_y'^2)\pm\sqrt{(\lambda_x'^2-\lambda_y'^2)^2+16}]$ for $\lambda_x'^2\lambda_y'^2>4$. Then Eq. (\ref{H3}) goes to
\begin{eqnarray}
\mathcal{H}_3&=&3\omega_r'd^\dagger d+\sum_{i=1}^2\Lambda_i\eta_i^\dagger\eta_i
-\frac{1}{2}g_x'(d+d^\dagger)[b_{x1}(\eta_1+\eta_1^\dagger)\nonumber\\
&&+b_{x2}(\eta_1+\eta_1^\dagger)\eta_2^\dagger\eta_2+b_{x3}\eta_1^\dagger\eta_1
(\eta_2+\eta_2^\dagger)]\nonumber\\
&&-\frac{1}{2}g_y'(d-d^\dagger)[b_{y1}(\eta_1^\dagger-\eta_1)+b_{y2}(\eta_1^\dagger-\eta_1)\eta_2^\dagger\eta_2\nonumber\\
&&+b_{y3}
\eta_1^\dagger\eta_1(\eta_2^\dagger-\eta_2)], \label{H32}
\end{eqnarray}
where $b_{x1}=-b_{x2}=b_{y3}=\xi_1$, $b_{x3}=b_{y1}=-b_{y2}=\xi_2$ for $\lambda_x'^2>\lambda_y'^2$ and $\lambda_x'^2\lambda_y'^2<4$, $b_{x1}=-b_{x2}=b_{y3}=-\xi_2$, $b_{x3}=b_{y1}=-b_{y2}=-\xi_1$ for $\lambda_x'^2<\lambda_y'^2$ and $\lambda_x'^2\lambda_y'^2<4$, $b_{x1}=b_{y1}=0$, $b_{x2}=b_{y3}=\xi_1$, $b_{x3}=-b_{y2}=-\xi_2$ for $\lambda_x'^2>\lambda_y'^2$ and $\lambda_x'^2\lambda_y'^2>4$, and $b_{x1}=b_{y1}=0$, $b_{x2}=-b_{y3}=-\xi_2$, $b_{x3}=-b_{y2}=-\xi_1$ for $\lambda_x'^2<\lambda_y'^2$ and $\lambda_x'^2\lambda_y'^2>4$. Here, $\xi_{1,2}=\sqrt{1+\frac{4}{\sqrt{(\lambda_x'^2-\lambda_y'^2)^2+16}}}\mp\sqrt{1-\frac{4}{\sqrt{(\lambda_x'^2-\lambda_y'^2)^2+16}}}$.

 In the $\omega_q'/\omega_r'\rightarrow\infty$ limit, we transform the Hamiltonian $\mathcal{H}_2$ with a unitary operator $U'=e^{S'}$. The detailed derivation is shown in Appendix \ref{appendixb}. Being projected to the low-energy subspace, the effective Hamiltonian is
\begin{widetext}
\begin{eqnarray}
\mathcal{H}_3^{\mathrm{eff}}&=&3\omega_r'd^\dagger d+\left\{
                               \begin{array}{ll}
                              -\frac{1}{4}g_{x}'^2\xi_1^2/\Lambda_1(d+d^\dagger)^2+\frac{1}{4}g_{y}'^2\xi_2^2/\Lambda_1(d-d^\dagger)^2, & \hbox{$ \lambda_x'^2>\lambda_y'^2 $ and $\lambda_x'^2\lambda_y'^2<4$,} \\
                                 -\frac{1}{4}g_{x}'^2\xi_2^2/\Lambda_1(d+d^\dagger)^2+\frac{1}{4}g_{y}'^2\xi_1^2/\Lambda_1(d-d^\dagger)^2, & \hbox{$\lambda_x'^2<\lambda_y'^2$ and $\lambda_x'^2\lambda_y'^2<4$,} \\
                                 0, & \hbox{$\lambda_x'^2\lambda_y'^2>4$.}
                               \end{array}
                             \right. \label{H3eff}
\end{eqnarray}
\end{widetext}
After diagonalizing the Hamiltonian, we obtain
\begin{widetext}
\begin{eqnarray}
\mathcal{H}_3^{\mathrm{eff}}&=&\varpi h^\dagger h
=\left\{\begin{array}{ll}
                              3\omega_r'\sqrt{(1-\frac{D_x\xi_1^2}{3\Lambda_1})(1-\frac{D_y\xi_2^2}{3\Lambda_1})}h^\dagger h, &\hbox{$ \lambda_x'^2>\lambda_y'^2 $ and $\lambda_x'^2\lambda_y'^2<4$,}\\
                               3\omega_r'\sqrt{(1-\frac{D_x\xi_2^2}{3\Lambda_1})(1-\frac{D_y\xi_1^2}{3\Lambda_1})}h^\dagger h, & \hbox{$\lambda_x'^2<\lambda_y'^2$ and $\lambda_x'^2\lambda_y'^2<4$,} \\
                                 3\omega_r'h^\dagger h, & \hbox{$\lambda_x'^2\lambda_y'^2>4$.}
                               \end{array}
\right.
\end{eqnarray}
\end{widetext}
Here, $\varpi$ is real and positive. Consequently, the phase transition is inhibited even in the ultrastrong-coupling regime.
\section{Three-Qubit model}
\begin{figure}
  \centering
  \includegraphics[width=0.4\textwidth]{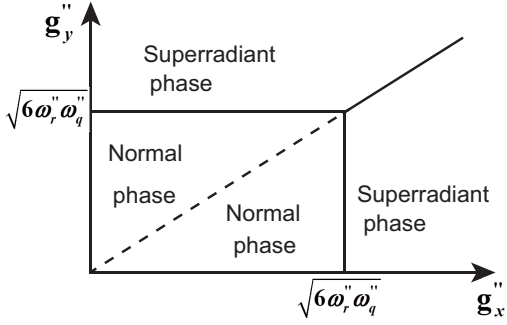}\\
  \caption{Phase diagram of three qubits model in the $(g_x'', g_y'')$ plane.}\label{fig4}
\end{figure}

The Hamiltonian for the three-qubit model is
\begin{eqnarray}
\mathcal{H}_4&=&3\omega_r''d'^\dagger d'+2\omega_q''S_z'-g_x''(d'+d'^\dagger)S_x'\nonumber\\
&&-ig_y''(d'-d'^\dagger)S_y'+D_x'S_x'^2+D_y'S_y'^2, \label{H4}
\end{eqnarray}
where $S_j(j=x,y,z)=\Sigma_{i=1}^{3}\frac{\sigma_i^j}{2}$. In the $\omega_q''/\omega_r''\rightarrow\infty$ limit, by applying second-order perturbation theory, we obtain an effective Hamiltonian, and the corresponding phase diagram is shown in Fig. \ref{fig4}. It is easy to see that the superradiant phase is restored.\\
\section{discussion}
Obviously, our results can be extended to the $N$-qubit situation for $g_x=g_y$. The Hamiltonian for the $N$-qubit model is
\begin{eqnarray}
\mathcal{H}_N&=&\mathcal{H}_{0}^{(N)}+\mathcal{V}^{(N)},\nonumber\\
\mathcal{H}_0^{(N)}&=&3\Omega_rt^\dagger t+2\Omega_qJ_z+DJ_x^2+DJ_y^2,\nonumber\\
\mathcal{V}^{(N)}&=&-g(t+t^\dagger)J_x-ig(t-t^\dagger)J_y.\label{hn}
\end{eqnarray}
The qubit Hamiltonian can be diagonalized as
\begin{eqnarray}
\mathcal{H}_{0q}^{(N)}&=&2\Omega_qJ_z+D(J^2-J_z^2)=\sum_{j,m_z}[2\Omega_qm_z\nonumber\\
&&+D (j(j+1)-m_z^2)]|j,m_z\rangle\langle j,m_z|,\label{h0q}
\end{eqnarray}
where $J$ is the total angular momentum of qubits, the spin number $j=0,1,2,...,\frac{N}{2}$($N$ is even) or $j=\frac{1}{2},\frac{3}{2},...,\frac{N}{2}$($N$ is odd) and the projection number $m_z=-j,-j+1,...,j$. The corresponding ground state is $|\frac{N}{2},-\frac{N}{2}\rangle$ for $D<2\Omega_q$. For $D>2\Omega_q$, the ground state is $|0,0\rangle$ for $N\in \mathrm{even}$ and $|\frac{1}{2},-\frac{1}{2}\rangle$ for $N\in \mathrm{odd}$. As shown in Appendix \ref{appendixc}, in the $\Omega_q/\Omega_r\rightarrow\infty$ limit, the effective Hamiltonian of Eq. (\ref{hn}) goes to
\begin{widetext}
\begin{eqnarray}
\mathcal{H}_{N}^{\mathrm{eff}}&=&\left\{\begin{array}{ll}
                              3\Omega_r(1-\frac{N\lambda^2}{3[(N-1)\lambda^2+2]})t^\dagger t, &\hbox{$\lambda^2<2$,} \\
                              3\Omega_rt^\dagger t , &\hbox{$\lambda^2>2$ and $N\in \mathrm{even}$,} \\
                                 3\Omega_r(1-\frac{\lambda^2}{6})t^\dagger t, &\hbox{$\lambda^2>2$ and $N\in \mathrm{odd}$.}
                               \end{array}
                             \right.\label{HN}
\end{eqnarray}
\end{widetext}
We can easily see that when $N$ is even, the phase transition is inhibited. However, when $N$ is odd, a superradiant phase transition occurs in the ultrastrong-coupling regime and the phase-transition point is at $\lambda=\sqrt{6}$.
\section{conclusion}
In summary, first, we made a fully quantum mechanical analysis of the one-qubit Hamiltonian by using an effective low-energy theory in the limit $\omega_q/\omega_r\rightarrow\infty$. A second-order phase-transition occurs and a Goldstone mode emerges in this limit. In addition, a first-order phase transition occurs between two different superradiant phases. Second, a two-qubit Hamiltonian beyond the Dicke model is analyzed fully quantum mechanically. We show that the quantum phase transition is inhibited even in the ultrastrong-coupling regime in this model. Third, a three-qubit model is analyzed and the QPT is restored. Finally, we extend our results to the $N$-qubit case.
\section*{ACKNOWLEDGMENTS}
 We acknowledge A. Li for helpful discussions. We acknowledge financial support in part by the 10000-Plan of Shandong Province (Taishan Scholars), NSFC Grant No. 11474182, Open Research Fund Program of the State Key Laboratory of Low-Dimensional Quantum Physics Grant No. KF201513, and the key R$\&$D Plan Project of Shandong Province, grant No. 2015GGX101035.\\

\begin{appendix}

\section{derivation of Eq. (\ref{Heff})\label{appendixa}}
The one-qubit Hamiltonian Eq. (\ref{Hamiltonian}) is
\begin{eqnarray}
\mathcal{H}&=&\mathcal{H}_0+\mathcal{V},\nonumber\\
\mathcal{H}_0&=&\omega_rb^\dagger b+\omega_q\frac{\sigma_z}{2},\nonumber\\
\mathcal{V}&=&-g_x(b+b^\dagger){\sigma_x\over2}-ig_y(b-b^\dagger){\sigma_y\over2}.\label{a1}
\end{eqnarray}
Here, $\mathcal{H}_0$ and $\mathcal{V}$ are block-diagonal and block-off-diagonal with respect to the spin subspace. We use an effective low-energy theory shown in Ref. \cite{flimit7} for a fully quantum mechanical analysis of the Hamiltonian. We consider a unitary transformation $U=e^S$ where $S$ is anti-Hermitian. The transformed Hamiltonian reads
\begin{eqnarray}
\mathcal{H}'&=&e^{S}\mathcal{H}e^{-S}=\mathcal{H}_0+\mathcal{V}+[S,\mathcal{H}_0+\mathcal{V}]\nonumber\\&&+\frac{1}{2!}[S,[S,\mathcal{H}_0
+\mathcal{V}]]+... .\label{a2}
\end{eqnarray}
We require that the block-off-diagonal terms maintain zero up to second order of $g_x$ and $g_y$, then
\begin{eqnarray}
[S_1,\mathcal{H}_0]=-\mathcal{V}.
\end{eqnarray}
With this requirement, we find the generator $S$ is
\begin{eqnarray}
S&=&S_1=\frac{g_x+g_y}{2\omega_q}(b^\dagger\sigma_{-}-b\sigma_{+})+\frac{g_x-g_y}{2\omega_q}(b\sigma_{-}-b^\dagger\sigma_{+})\nonumber\\
&&+\textit{O}((g_x,g_y)\frac{\omega_r}{\omega_q^2}).
\end{eqnarray}
By projecting Hamiltonian Eq. (\ref{a2}) to the low-energy spin subspace, we obtain
\begin{eqnarray}
\mathcal{H}_{\mathrm{eff}}&=&P\mathcal{H}'P\nonumber\\
&=&P(\mathcal{H}_0+\frac{1}{2}[S_1, \mathcal{V}])P\nonumber\\
&=&\omega_rb^\dagger b-\omega_r\frac{\lambda_x^2}{4}(b+b^\dagger)^2+\omega_r\frac{\lambda_y^2}{4}(b-b^\dagger)^2\nonumber\\
&&-\frac{\omega_q}{2}.
\end{eqnarray}
\section{derivation of Eq. (\ref{H3eff})\label{appendixb}}
The two-qubit Hamiltonian Eq. (\ref{H32}) is
\begin{eqnarray}
\mathcal{H}_3&=&\mathcal{H}_{30}+\mathcal{V}_{3},\nonumber\\
\mathcal{H}_{30}&=&3\omega_r'd^\dagger d+\sum_{i=1}^2\Lambda_i\eta_i^\dagger\eta_i,\nonumber\\
\mathcal{V}_{3}&=&-\frac{1}{2}g_x'(d+d^\dagger)[b_{x1}(\eta_1+\eta_1^\dagger)\nonumber\\
&&+b_{x2}(\eta_1+\eta_1^\dagger)\eta_2^\dagger\eta_2+b_{x3}\eta_1^\dagger\eta_1
(\eta_2+\eta_2^\dagger)]\nonumber\\
&&-\frac{1}{2}g_y'(d-d^\dagger)[b_{y1}(\eta_1^\dagger-\eta_1)\nonumber\\
&&+b_{y2}(\eta_1^\dagger-\eta_1)\eta_2^\dagger\eta_2
+b_{y3}
\eta_1^\dagger\eta_1(\eta_2^\dagger-\eta_2)],\label{b1}
\end{eqnarray}
where Hamiltonians $\mathcal{H}_{30}$ and $\mathcal{V}_{3}$ are block-diagonal and block-off-diagonal with respect to the  spin subspace. We again consider a unitary transformation $U'=e^{S'}$ where $S'$ is anti-Hermitian. The transformed Hamiltonian reads
\begin{eqnarray}
\mathcal{H}_{3}'&=&e^{S'}\mathcal{H}_3e^{-S'}=\mathcal{H}_{30}+\mathcal{V}_{3}+[S',\mathcal{H}_{30}+\mathcal{V}_3]\nonumber\\
&&+\frac{1}{2!}[S',[S',\mathcal{H}_{30}
+\mathcal{V}_3]]+... . \label{b2}
\end{eqnarray}
With the requirement that block-off-diagonal terms maintain zero up to second order of $g_x'$ and $g_y'$, then
\begin{eqnarray}
[S_1',\mathcal{H}_{30}]&=&-\mathcal{V}_3\nonumber\\
\mathcal{H}_{32}^{d}+\mathcal{H}_{32}^{od}&=&[S_1',\mathcal{V}_3]+\frac{1}{2!}[S_1',[S_1',\mathcal{H}_{30}]],\nonumber\\
\mathcal{H}_{32}^{od}&=&-[S_2',\mathcal{H}_{30}],
\end{eqnarray}
where $\mathcal{H}_{32}^{d}$ and $\mathcal{H}_{32}^{od}$ are diagonal and off-diagonal with respect to the spin subspace. By projecting Hamiltonian Eq. (\ref{b2}) to the low-energy spin subspace, we obtain
\begin{widetext}
\begin{eqnarray}
\mathcal{H}_3^{\mathrm{eff}}&=&P\mathcal{H}_3P\nonumber\\
&=&P(\mathcal{H}_{30}+\frac{1}{2}[S_1', \mathcal{V}_3]+[S_2',\mathcal{H}_{30}])P\nonumber\\
&=&3\omega_r'd^\dagger d+\left\{
                               \begin{array}{ll}
                              -\frac{1}{4}g_{x}'^2\xi_1^2/\Lambda_1(d+d^\dagger)^2+\frac{1}{4}g_{y}'^2\xi_2^2/\Lambda_1(d-d^\dagger)^2, &\begin{array}{ll} \hbox{$ \lambda_x'^2>\lambda_y'^2 $ and $\lambda_x'^2\lambda_y'^2<4$,}\end{array} \\
                                 -\frac{1}{4}g_{x}'^2\xi_2^2/\Lambda_1(d+d^\dagger)^2+\frac{1}{4}g_{y}'^2\xi_1^2/\Lambda_1(d-d^\dagger)^2, & \begin{array}{ll}\hbox{$\lambda_x'^2<\lambda_y'^2$   and $\lambda_x'^2\lambda_y'^2<4$,}\end{array} \\
                                 0, &\hbox{$\lambda_x'^2\lambda_y'^2>4$.}
                               \end{array}
                             \right.
\end{eqnarray}
\end{widetext}
\section{derivation of Eq. (\ref{HN} \label{appendixc})}
The $N$-qubit Hamiltonian Eq. (\ref{hn}) is
\begin{eqnarray}
\mathcal{H}_N&=&\mathcal{H}_{0}^{(N)}+\mathcal{V}^{(N)},\nonumber\\
\mathcal{H}_0^{(N)}&=&3\Omega_rt^\dagger t+2\Omega_qJ_z+DJ_x^2+DJ_y^2,\nonumber\\
\mathcal{V}^{(N)}&=&-g(t+t^\dagger)J_x-ig(t-t^\dagger)J_y,
\end{eqnarray}
where $\mathcal{H}_{0q}^{(N)}$ can be diagonalized as shown in Eq. (\ref{h0q}). Thus Hamiltonians $\mathcal{H}_{0}^{(N)}$ and $\mathcal{V}^{(N)}$ are block-diagonal and block-off-diagonal with respect to the  spin subspace. We make a unitary transformation $U^{(N)}=e^{S^{(N)}}$ of the Hamiltonian $\mathcal{H}_N$ where $S^{(N)}$ is anti-Hermitian. Then,
\begin{eqnarray}
\mathcal{H}_{N}'&=&e^{S^{(N)}}\mathcal{H}_Ne^{-S^{(N)}}=\mathcal{H}_{0}^{(N)}+\mathcal{V}^{(N)}
+[S^{(N)},\mathcal{H}_{0}^{(N)}+\nonumber\\&&\mathcal{V}^{(N)}]
+\frac{1}{2!}[S^{(N)},[S^{(N)},\mathcal{H}_{0}^{(N)}
+\mathcal{V}^{(N)}]]+... . \label{c2}
\end{eqnarray}
With the requirement that block-off-diagonal terms maintain zero up to second order of $g$, then
\begin{eqnarray}
[S_1^{(N)},\mathcal{H}_{0}^{(N)}]&=&-\mathcal{V}^{(N)},
\end{eqnarray}
with
\begin{eqnarray}
S_{1}^{(N)}&=&\sum_{j,m_z}a_{j,m_z}t|j,m_z+1\rangle\langle j,m_z|\nonumber\\
&&+b_{j,m_z}t^\dagger|j,m_z-1\rangle\langle j,m_z|,
\end{eqnarray}
where $a_{j,m_z}=g\frac{\sqrt{j(j+1)-m_z(m_z+1)}}{-2\Omega_q+D(1+2m_z)}$ and $b_{j,m_z}=g\frac{\sqrt{j(j+1)-m_z(m_z-1)}}{2\Omega_q+D(1-2m_z)}$. After projecting Hamiltonian Eq. (\ref{c2}) to the low-energy subspace, Eq. (\ref{HN}) is obtained that
\begin{widetext}
\begin{eqnarray}
\mathcal{H}_{N}^{\mathrm{eff}}=P\mathcal{H}_{N}'P&=&\left\{\begin{array}{ll}
                              3\Omega_r(1-\frac{N\lambda^2}{3[(N-1)\lambda^2+2]})t^\dagger t, &\hbox{$\lambda^2<2$,} \\
                              3\Omega_rt^\dagger t , &\hbox{$\lambda^2>2$ and $N\in \mathrm{even}$,} \\
                                 3\Omega_r(1-\frac{\lambda^2}{6})t^\dagger t, &\hbox{$\lambda^2>2$ and $N\in \mathrm{odd}$.}
                               \end{array}
                             \right.
\end{eqnarray}
\end{widetext}
  \end{appendix}

\end{document}